\documentstyle[12pt]{article}

\catcode`\@=11
\long\def\@makefntext#1{
\protect\noindent \hbox to 3.2pt {\hskip-.9pt
$^{{\ninerm\@thefnmark}}$\hfil}#1\hfill}		

\def\@makefnmark{\hbox to 0pt{$^{\@thefnmark}$\hss}}  

\def\ps@myheadings{\let\@mkboth\@gobbletwo
\def\@oddhead{\hbox{}
\rightmark\hfil\ninerm\thepage}
\def\@oddfoot{}\def\@evenhead{\ninerm\thepage\hfil
\leftmark\hbox{}}\def\@evenfoot{}
\def\sectionmark##1{}\def\subsectionmark##1{}}

\newcounter{sectionc}\newcounter{subsectionc}\newcounter{subsubsectionc}
\renewcommand{\section}[1] {\vspace*{0.6cm}\addtocounter{sectionc}{1}
\setcounter{subsectionc}{0}\setcounter{subsubsectionc}{0}\noindent
	{\normalsize\bf\thesectionc. #1}\par\vspace*{0.4cm}}
\renewcommand{\subsection}[1] {\vspace*{0.6cm}\addtocounter{subsectionc}{1}
	\setcounter{subsubsectionc}{0}\noindent
	{\normalsize\it\thesectionc.\thesubsectionc. #1}\par\vspace*{0.4cm}}
\renewcommand{\subsubsection}[1]
{\vspace*{0.6cm}\addtocounter{subsubsectionc}{1}
	\noindent {\normalsize\rm\thesectionc.\thesubsectionc.\thesubsubsectionc.
	#1}\par\vspace*{0.4cm}}

\newcounter{appendixc}
\newcounter{subappendixc}[appendixc]
\newcounter{subsubappendixc}[subappendixc]

\renewcommand{\appendix}[1] {\vspace*{0.6cm}
        \refstepcounter{appendixc}
        \setcounter{figure}{0}
        \setcounter{table}{0}
        \setcounter{equation}{0}
        \renewcommand{\thefigure}{\Alph{appendixc}.\arabic{figure}}
        \renewcommand{\thetable}{\Alph{appendixc}.\arabic{table}}
        \renewcommand{\theappendixc}{\Alph{appendixc}}
        \renewcommand{\theequation}{\Alph{appendixc}.\arabic{equation}}
        \noindent{\bf Appendix \theappendixc #1}\par\vspace*{0.4cm}}

\def\abstracts#1{{

\centering{\begin{minipage}{12.2truecm}\footnotesize\baselineskip=12pt\noindent
	\centerline{\footnotesize ABSTRACT}\vspace*{0.3cm}
	\parindent=0pt #1
	\end{minipage}}\par}}

\renewenvironment{thebibliography}[1]
	{\begin{list}{\arabic{enumi}.}
	{\usecounter{enumi}\setlength{\parsep}{0pt}
\setlength{\leftmargin 1.25cm}{\rightmargin 0pt}
	 \setlength{\itemsep}{0pt} \settowidth
	{\labelwidth}{#1.}\sloppy}}{\end{list}}

\newcounter{itemlistc}
\newcounter{romanlistc}
\newcounter{alphlistc}
\newcounter{arabiclistc}

\newcommand{\fcaption}[1]{
        \refstepcounter{figure}
        \setbox\@tempboxa = \hbox{\footnotesize Fig.~\thefigure. #1}
        \ifdim \wd\@tempboxa > 6in
           {\begin{center}
        \parbox{6in}{\footnotesize\baselineskip=12pt Fig.~\thefigure. #1}
            \end{center}}
        \else
             {\begin{center}
             {\footnotesize Fig.~\thefigure. #1}
              \end{center}}
        \fi}

\newcommand{\tcaption}[1]{
        \refstepcounter{table}
        \setbox\@tempboxa = \hbox{\footnotesize Table~\thetable. #1}
        \ifdim \wd\@tempboxa > 6in
           {\begin{center}
        \parbox{6in}{\footnotesize\baselineskip=12pt Table~\thetable. #1}
            \end{center}}
        \else
             {\begin{center}
             {\footnotesize Table~\thetable. #1}
              \end{center}}
        \fi}

\def\@citex[#1]#2{\if@filesw\immediate\write\@auxout
	{\string\citation{#2}}\fi
\def\@citea{}\@cite{\@for\@citeb:=#2\do
	{\@citea\def\@citea{,}\@ifundefined
	{b@\@citeb}{{\bf ?}\@warning
	{Citation `\@citeb' on page \thepage \space undefined}}
	{\csname b@\@citeb\endcsname}}}{#1}}

\newif\if@cghi
\def\cite{\@cghitrue\@ifnextchar [{\@tempswatrue
	\@citex}{\@tempswafalse\@citex[]}}
\def\citelow{\@cghifalse\@ifnextchar [{\@tempswatrue
	\@citex}{\@tempswafalse\@citex[]}}
\def\@cite#1#2{{$\null^{#1}$\if@tempswa\typeout
	{IJCGA warning: optional citation argument
	ignored: `#2'} \fi}}

 1
 1
 1

\font\ninerm=cmr9

\begin{document}

\centerline{\normalsize\bf GREENSITE-HALPERN STABILIZATION OF $A_k$}
\baselineskip=16pt
\centerline{\normalsize\bf SINGULARITIES IN THE $N \rightarrow \infty$ LIMIT}
\baselineskip=16pt

\vspace*{0.6cm}
\centerline{\footnotesize J. MAEDER and W. R\"UHL}
\baselineskip=13pt
\centerline{\footnotesize\it Department of Physics, University of
Kaiserslautern, P.O.Box 3049}
\centerline{\footnotesize\it 67653 Kaiserslautern, Germany}
\centerline{\footnotesize E-mail: ruehl@gypsy.physik.uni-kl.de}
\vspace*{0.3cm}

\vspace*{0.9cm}
\abstracts{The Greensite-Halpern method of stabilizing bottomless Euclidean
actions is applied to zerodimensional O(N) sigma models with unstable
$A_k$ singularities in the $N = \infty$ limit.
\\*[1cm]
Dedicated to J. Lukierski to his 60th birthday.}
\vspace*{1.5cm}

\message{reelletc.tex (Version 1.0): Befehle zur Darstellung |R  |N, Aufruf
z.B. \string\bbbr}
%
%
\message{reelletc.tex (Version 1.0): Befehle zur Darstellung |R  |N, Aufruf
z.B. \string\bbbr}
%
%
%
%
%
\font \smallescriptscriptfont = cmr5
\font \smallescriptfont       = cmr5 at 7pt
\font \smalletextfont         = cmr5 at 10pt
\font \tensans                = cmss10
\font \fivesans               = cmss10 at 5pt
\font \sixsans                = cmss10 at 6pt
\font \sevensans              = cmss10 at 7pt
\font \ninesans               = cmss10 at 9pt
\newfam\sansfam
\textfont\sansfam=\tensans\scriptfont\sansfam=\sevensans
\scriptscriptfont\sansfam=\fivesans
\def\sans{\fam\sansfam\tensans}
\def\bbbr{{\rm I\!R}} 
\def\bbbn{{\rm I\!N}} 
\def\bbbE{{\rm I\!E}} 
\def\bbbm{{\rm I\!M}}
\def\bbbh{{\rm I\!H}}
\def\bbbk{{\rm I\!K}}
\def\bbbd{{\rm I\!D}}
\def\bbbp{{\rm I\!P}}
\def\bbbone{{\mathchoice {\rm 1\mskip-4mu l} {\rm 1\mskip-4mu l}
{\rm 1\mskip-4.5mu l} {\rm 1\mskip-5mu l}}}
\def\bbbc{{\mathchoice {\setbox0=\hbox{$\displaystyle\rm C$}\hbox{\hbox
to0pt{\kern0.4\wd0\vrule height0.9\ht0\hss}\box0}}
{\setbox0=\hbox{$\textstyle\rm C$}\hbox{\hbox
to0pt{\kern0.4\wd0\vrule height0.9\ht0\hss}\box0}}
{\setbox0=\hbox{$\scriptstyle\rm C$}\hbox{\hbox
to0pt{\kern0.4\wd0\vrule height0.9\ht0\hss}\box0}}
{\setbox0=\hbox{$\scriptscriptstyle\rm C$}\hbox{\hbox
to0pt{\kern0.4\wd0\vrule height0.9\ht0\hss}\box0}}}}

\def\bbbe{{\mathchoice {\setbox0=\hbox{\smalletextfont e}\hbox{\raise
0.1\ht0\hbox to0pt{\kern0.4\wd0\vrule width0.3pt height0.7\ht0\hss}\box0}}
{\setbox0=\hbox{\smalletextfont e}\hbox{\raise
0.1\ht0\hbox to0pt{\kern0.4\wd0\vrule width0.3pt height0.7\ht0\hss}\box0}}
{\setbox0=\hbox{\smallescriptfont e}\hbox{\raise
0.1\ht0\hbox to0pt{\kern0.5\wd0\vrule width0.2pt height0.7\ht0\hss}\box0}}
{\setbox0=\hbox{\smallescriptscriptfont e}\hbox{\raise
0.1\ht0\hbox to0pt{\kern0.4\wd0\vrule width0.2pt height0.7\ht0\hss}\box0}}}}

\def\bbbq{{\mathchoice {\setbox0=\hbox{$\displaystyle\rm Q$}\hbox{\raise
0.15\ht0\hbox to0pt{\kern0.4\wd0\vrule height0.8\ht0\hss}\box0}}
{\setbox0=\hbox{$\textstyle\rm Q$}\hbox{\raise
0.15\ht0\hbox to0pt{\kern0.4\wd0\vrule height0.8\ht0\hss}\box0}}
{\setbox0=\hbox{$\scriptstyle\rm Q$}\hbox{\raise
0.15\ht0\hbox to0pt{\kern0.4\wd0\vrule height0.7\ht0\hss}\box0}}
{\setbox0=\hbox{$\scriptscriptstyle\rm Q$}\hbox{\raise
0.15\ht0\hbox to0pt{\kern0.4\wd0\vrule height0.7\ht0\hss}\box0}}}}

\def\bbbt{{\mathchoice {\setbox0=\hbox{$\displaystyle\rm
T$}\hbox{\hbox to0pt{\kern0.3\wd0\vrule height0.9\ht0\hss}\box0}}
{\setbox0=\hbox{$\textstyle\rm T$}\hbox{\hbox
to0pt{\kern0.3\wd0\vrule height0.9\ht0\hss}\box0}}
{\setbox0=\hbox{$\scriptstyle\rm T$}\hbox{\hbox
to0pt{\kern0.3\wd0\vrule height0.9\ht0\hss}\box0}}
{\setbox0=\hbox{$\scriptscriptstyle\rm T$}\hbox{\hbox
to0pt{\kern0.3\wd0\vrule height0.9\ht0\hss}\box0}}}}

\def\bbbs{{\mathchoice
{\setbox0=\hbox{$\displaystyle     \rm S$}\hbox{\raise0.5\ht0\hbox
to0pt{\kern0.35\wd0\vrule height0.45\ht0\hss}\hbox
to0pt{\kern0.55\wd0\vrule height0.5\ht0\hss}\box0}}
{\setbox0=\hbox{$\textstyle        \rm S$}\hbox{\raise0.5\ht0\hbox
to0pt{\kern0.35\wd0\vrule height0.45\ht0\hss}\hbox
to0pt{\kern0.55\wd0\vrule height0.5\ht0\hss}\box0}}
{\setbox0=\hbox{$\scriptstyle      \rm S$}\hbox{\raise0.5\ht0\hbox
to0pt{\kern0.35\wd0\vrule height0.45\ht0\hss}\raise0.05\ht0\hbox
to0pt{\kern0.5\wd0\vrule height0.45\ht0\hss}\box0}}
{\setbox0=\hbox{$\scriptscriptstyle\rm S$}\hbox{\raise0.5\ht0\hbox
to0pt{\kern0.4\wd0\vrule height0.45\ht0\hss}\raise0.05\ht0\hbox
to0pt{\kern0.55\wd0\vrule height0.45\ht0\hss}\box0}}}}

\def\bbbz{{\mathchoice {\hbox{$\sans\textstyle Z\kern-0.4em Z$}}
{\hbox{$\sans\textstyle Z\kern-0.4em Z$}}
{\hbox{$\sans\scriptstyle Z\kern-0.3em Z$}}
{\hbox{$\sans\scriptscriptstyle Z\kern-0.2em Z$}}}}
%
%

 \setlength{\baselineskip}{15pt}

\noindent
1. Classical actions which are unbounded from below do not define Euclidean
quantum field theories because the partition functions diverge. A method
to modify the classical actions in such a fashion that convergence is
guaranteed on the one hand whereas the classical actions are only minimally
changed on the other hand has been proposed by Greensite and Halpern
\cite{1}. We refer to this method as "Greensite-Halpern stabilization".
Modifications of a theory are considered minimal if the stabilized and the
original "bottomless" theory have the same
\begin{enumerate}
\item classical limit;
\item perturbative series;
\item $N \rightarrow \infty$ limit.
\end{enumerate}
In \cite{1} it has been proved for typical models that these requirements
are indeed fulfilled. The Greensite-Halpern stabilization applied to a
stable theory leaves it unchanged.

A famous example of a classical bottomless theory is Euclidean gravity.
The same problem of instability arises in matrix models of pure
gravity. Applications of Greensite-Halpern stabilization to these models
can be found in \cite{2,3}.

The most popular method of stabilization is analytic continuation of the
classical action in a coupling constant. Expectation values are then not
necessarily analytic \cite{4} but it seems that the perturbative series
is always invariant under continuation. So the three axioms of minimality
formulated by Greensite and Halpern may also be fulfilled. It is, however,
known that both stabilization methods are inequivalent.

We want to apply the Greensite-Halpern stabilization method to zero
dimensional sigma models that exhibit $A_k$ singularities with $k > 1$,
($k = 1$ appears in \cite{1}). In these cases we have to perform double
scaling limits, where $N$ goes to infinity and coupling constants $\{f_r\}$
tend to their critical values $\{f^c_r\}$. There arise scale invariant
variables $\{\zeta_r\}^{k-1}_1$ and the singular factor in the partition
function is a generalized Airy function depending on these variables (see
\cite{5} for the details). The cases $A_k$ with $k = 2n$ are unstable.
If $k = 2n+1$ there are two signs $A_{2n+1}^{\pm}$ one of which (the
"wrong sign" $A^-_{2n+1}$) is also unstable.

The generalized Airy functions are given by integral representations. In the
stable cases the integral contours are the real axis. Mathematical textbooks
\cite{6} teach us that we have to choose complex contours in the unstable
cases. Though this leads to well-defined Airy functions, it is not clear
whether they are suited for a probabilistic interpretation in at least a
subdomain of the variables $\{\zeta_r\}$. At the end of this article we will
make a clarifying comment on this problem. On the other hand the
Greensite-Halpern stabilized theories have an obvious probabilistic
interpretation for all $\{\zeta_r\} \in \bbbr_{k-1}$.

\vspace{0.5cm}
\noindent
2. We consider zerodimensional sigma models
\begin{equation}
Z = \int \prod^N_{a=1} d\phi_a e^{-S}, \quad \phi_a \in \bbbr
\label{1}
\end{equation}
\begin{equation}
S = \frac12 \phi \cdot \phi + \sum^k_{r=2} \frac{f_r}{2r} N^{-r+1}
(\phi \cdot \phi)^r
\label{2}
\end{equation}
\begin{equation}
\phi \cdot \phi = Nz
\label{3}
\end{equation}
\begin{equation}
\tilde{S}(z) = \frac1N S = \frac12 z + \sum^k_{r=2} \frac{f_r}{2r} z^r.
\label{4}
\end{equation}
Angular integration gives
\begin{equation}
Z = \frac{(\pi N)^{\frac N2}}{\Gamma(\frac{N}{2})} \int^{\infty}_0 dz \frac1z
\exp N (\frac12 \log z - \tilde{S}(z)).
\label{5}
\end{equation}
The exponent in (\ref{5}) may exhibit a singularity $A_n (n \le k)$ which in
the limit $N \rightarrow \infty$ allows us to expand $Z$ and any expectation
value in a series of fractional negative powers of $N$. In the present context
we will deal with only the leading term which for $Z$ gives a generalized
Airy function.

The starting point of the Greensite-Halpern stabilization is the
Schr\"odinger-
equation
\begin{equation}
\left[ - \frac12 \Delta_{\phi} + \frac18 \sum_{a} \left( \frac{\partial
S}{\partial \phi_a}
 \right)^{2} - \frac14 \Delta_{\phi}S \right] \psi_0(\phi) = E_0\psi_0(\phi)
\label{6}
\end{equation}
with normalized ground state wave function $\psi_0(\phi)$ and eigenvalue
$E_0$. The ill-defined probability density
\[
\frac1Z e^{-S}
\]
is replaced by $|\psi_0(\phi)|^2$. Change of the coordinates (\ref{3})
and action (\ref{4}) gives
\begin{equation}
\left[ - \frac{2z}{N} \frac{\partial^2}{\partial z^2} -
\frac{\partial}{\partial z}
+ NV(Z)\right] \tilde{\psi}_0(z) = E_0 \tilde{\psi}_0(z)
\label{7}
\end{equation}
\begin{equation}
\psi_0(\phi) = \tilde{\psi}_0(z(\phi))
\label{8}
\end{equation}
\begin{equation}
 V(z) = \frac12 z (\tilde{S}^{\prime})^2 - \frac12 \tilde{S}^{\prime}
- \frac zN \tilde{S}^{\prime\prime}.
\label{9}
\end{equation}
Next we apply the $N \rightarrow \infty$ limit to the equation (\ref{7})
\cite{1,7}: We factorize
\begin{equation}
\tilde{\psi}_0(z) = z^{- \frac N4} \varphi_0(z)
\label{10}
\end{equation}
and rescale the equation
\begin{equation}
\left[ - \frac{2z}{N} \frac{\partial^2}{\partial z^2} + N\left(\frac{1}{8z}
+V(z)\right) + O(1) \right] \varphi_0(z) = E_0\varphi_0(z)
\label{11}
\end{equation}
in the neighborhood of the singularity.

If this singularity is $A_1$, its location $z_0$ is determined from
\begin{equation}
- \frac{1}{8z^2_0} + V^{\prime}(z_0) = 0.
\label{12}
\end{equation}
It has been shown in \cite{1} that the left hand side factorizes
\begin{equation}
- \frac{1}{8 z^2} + V^{\prime}(z) = F_1(z)F_2(z)
\label{13}
\end{equation}
with
\begin{equation}
F_1(z) = \tilde{S}^{\prime}(z) - \frac{1}{2z}
\label{14}
\end{equation}
\begin{equation}
F_2(z) = z \tilde{S}^{\prime\prime}(z) + \frac12 \tilde{S}^{\prime}(z)
+ \frac{1}{4z}.
\label{15}
\end{equation}
If
\begin{equation}
F_1(z_0) = 0
\label{16}
\end{equation}
we have an $A_1$ singularity in the action (\ref{5}) as well. An additional
branch of $A_1$ singularities in the potential of the Schr\"odinger
equation (\ref{11}) arises at
\begin{equation}
F_2(z_0) = 0.
\label{17}
\end{equation}
We will not consider it here (see, however, \cite{1}). The ground state
energy is in this approximation
\begin{eqnarray}
E_0 &=& N\left(\frac{1}{8z_0} + V(z_0)\right) \nonumber \\
&=& \frac12 Nz_0 \left(\tilde{S}^{\prime}(z_0)\right)^2 \ge 0
\label{18}
\end{eqnarray}
which remains valid in the $A_k, k > 1$, case.

To complete the discussion of the $A_1$ case we prove stability (i.e.
$A_1$ is $A^+_1$). We expand the potential to next order
\begin{equation}
\frac{1}{8z} + V(z) = \frac{1}{8z_0} + V(z_0) + \frac12(z-z_0)^2
\omega^2 + O((z-z_0)^3)
\label{19}
\end{equation}
and
\begin{eqnarray}
\omega^2 &=& F^{\prime}_1(z_0) F_2(z_0) + F_1(z_0)F^{\prime}_2(z_0) \nonumber
\\
&=& F_1^{\prime}(z_0)F_2(z_0)
\label{20}
\end{eqnarray}
if (\ref{16}) holds.

Now from (\ref{14}), (\ref{15}) we obtain
\begin{equation}
F_2(z) = zF^{\prime}_1(z) + \frac12 F_1(z)
\label{21}
\end{equation}
so that once again from (\ref{16})
\begin{equation}
\omega^2 = z_0(F^{\prime}_1(z_0))^2 > 0
\label{22}
\end{equation}
and we have (local) stability. We will later see that any
$A_{n+1}$ singularity in the action (\ref{5}) implies a (stable) $A_{n+1}$
singularity in the potential of the Schr\"odinger equation. Other
singularities in the potential (such as $A_1$ (\ref{17})) are not
automatically stable.

The ground state energy $E_0$ is to next order
\begin{equation}
E_0 = N\left(\frac{1}{8z_0}+ V(z_0)\right) + \epsilon_1
\label{23}
\end{equation}
where to leading order now
\begin{equation}
\left[- \frac{2z_0}{N} \frac{\partial^2}{\partial z^2} + \frac12
N \omega^2(z-z_0)^2\right] \varphi_0(z) = \epsilon_1\varphi_0(z).
\label{24}
\end{equation}
This equation is rescaled by
\begin{equation}
x = N^{\frac12} (z-z_0)
\label{25}
\end{equation}
so that the oscillator equation
\begin{equation}
\left[ - \frac12 \frac{\partial^2}{\partial x^2} + \frac12 \frac{\omega^2}
{4z_0}x^2\right] \varphi_0(z(x)) = \frac{\epsilon_1}{4z_0}
\varphi_0(z(x))
\label{26}
\end{equation}
results. It follows
\begin{equation}
\epsilon_1 = z_0^{\frac12}\omega
\label{27}
\end{equation}
and
\begin{equation}
\varphi_0(z(x)) = A \cdot e^{- \frac12 \frac{\omega}{\sqrt{ \mbox{\scriptsize
$4 z_0$}}} x^2}.
\label{28}
\end{equation}

\vspace{0,5cm}
\noindent
3. An $A_{n+1}$ singularity in (\ref{5}) shows up at $z_0$ if
\begin{eqnarray}
F_1^{(m)}(z_0) = 0, \quad 0 \le m \le n \nonumber \\
F_1^{(n+1)}(z_0) \neq 0.
\label{29}
\end{eqnarray}
If in (\ref{2}) and (\ref{4}) we choose $k = n+1$ (the "minimal
set" of coupling constants) there is exactly one such singularity and
corresponding critical coupling constants $\{f^c_r\}^{n+1}_2$ (see \cite{5}).
Since from (\ref{21})
\begin{equation}
F_2^{(m)} = zF_1^{(m+1)} + (m + \frac12)F_1^{(m)}
\label{30}
\end{equation}
(\ref{29}) implies
\begin{eqnarray}
F_2^{(m)}(z_0) = 0, \quad 0 \le m \le n-1 \nonumber \\
F_2^{(n)}(z_0) \neq 0.
\label{31}
\end{eqnarray}
At such point $z_0$
\begin{equation}
(F_1(z)F_2(z))^{(2n+1)}|_{z_0} = \left(2n+1 \atop n+1 \right) z_0
\left( F_1^{(n+1)}(z_0)\right)^2 > 0
\label{32}
\end{equation}
whereas
\begin{equation}
(F_1(z)F_2(z))^{(m)}|_{z_0} = 0, \quad m \le 2n.
\label{33}
\end{equation}
It follows that at leading order in $z-z_0$ the potential in the Schr\"odinger
equation is
\begin{equation}
+ N \cdot \frac{g_{2n+2}}{2n+2} (z-z_0)^{2n+2}
\label{34}
\end{equation}
with
\begin{equation}
g_{2n+2} = \frac{z_0(F_1^{(n+1)}(z_0))^2}{n!(n+1)!} > 0.
\label{35}
\end{equation}
So the Greensite-Halpern program produces a stable potential in the
Schr\"odinger equation for each $A_{n+1}$.

If the Schr\"odinger equation is rescaled at $N \rightarrow \infty$ in
analogy to (\ref{24}), (\ref{25}) we obtain
\begin{eqnarray}
\left( - \frac12 \frac{\partial^2}{\partial x^2} + \frac12 \frac{1}{n+1}
x^{2n+2}\right) \varphi_0(z(x)) = \nonumber \\
= \frac{N}{\lambda^2} \cdot \frac{\epsilon_1}{4z_0} \cdot \varphi_0(z(x))
\label{36}
\end{eqnarray}
where
\begin{equation}
x = \lambda(z-z_0)
\label{37}
\end{equation}
and
\begin{equation}
\lambda = \left( \frac{N^2g_{2n+2}}{4z_0}\right) ^{\frac{1}{2n+4}}.
\label{38}
\end{equation}
So $\varphi_0(z(x))$ is a universal function of $x$ and
\begin{equation}
\epsilon_1 = \frac{4z_0\lambda^2}{N} \tilde{\epsilon}_1
\label{39}
\end{equation}
where $\tilde{\epsilon}_1$ is a universal number (depending on $n$). The
function
$\varphi_0(z(x)) = \chi_0(x)$ is symmetric in $x$ and for $x \rightarrow
\infty$ behaves as $(n > 0)$
\begin{eqnarray}
\chi_0(x) = A \exp \Big\{&-& \frac{x^{n+2}}{(n+1)^{\frac12}(n+2)} +
\frac12 (n+1) \log x \nonumber \\
&-& \frac{(n+1)^{\frac12}}{n} \tilde{\epsilon}_1 x^{-n} + O(x^{-n-2}) \Big\}.
\label{40}
\end{eqnarray}
Squaring this function and substituting (\ref{37}), (\ref{38}) we obtain
the Greensite-Halpern probability distribution over the real z-axis for
large $|z|$
\begin{equation}
|\varphi_0(z)|^2 = A^2 \exp \left\{- N \frac{|F_1^{(n+1)}(z_0)|}
{(n+2)!} |z-z_0|^{n+2} + O(\log z)\right\}
\label{41}
\end{equation}
(and analogously for large $N$). Here the effect of the stabilization can
be clearly seen: all "wrong signs"are eliminated.

Now we consider a deformed $A_{n+1}$ singularity: the coupling constants
$\{f_r\}$ are different from the critical ones $\{f^c_r\}$
\begin{equation}
f_r - f^c_r = \Theta_r
\label{42}
\end{equation}
but with $N \rightarrow \infty$ these $\Theta_r$ go to zero in such a
fashion that
\begin{eqnarray}
G(x;\{\zeta\}) &=& \lim_{N \to \infty} N \left\{\tilde{S}(z) -
\tilde{S}(z_0) - \frac12 \log \frac{z}{z_0}\right\} \nonumber \\
&=& \sum^n_{r=1} \frac{\zeta_r}{r!} x^r + \epsilon \frac{x^{n+2}}{(n+2)!}
\label{43}
\end{eqnarray}
($\epsilon = \pm1$ for even n).

\noindent
Thus in terms of $\lambda$ (\ref{37}) (the normalization in (\ref{38}) is
marginally changed)
\begin{equation}
\zeta_r = \lim_{N \to\infty} N \lambda^{-r} F_1^{(r-1)}(z_0)
\label{44}
\end{equation}
and the point $z_0$ is kept fixed by the requirement that the power of
order $n+1$ in (\ref{43}) vanishes. In this case the saddle point
integration of (\ref{5}) gives
\begin{equation}
Z_{\mbox{\scriptsize sing}} = \frac{(\pi N)^{\frac N2}}{\Gamma(\frac N2)}
\frac{1}{\lambda z_0}
\int_C dx \exp\{- G(x;\{\zeta\})\}
\label{45}
\end{equation}
where $C$ is a chain running from infinity to infinity along which the
integral converges exponentially. The integral is a generalized Airy
function.

Now we apply the analogous procedure in the Greensite-Halpern stabilization
program, which results in a measure
\begin{equation}
\frac{d\mu_{GH}}{dx} = \exp\{- \tilde{G}(x;\{\zeta\})\}
\label{46}
\end{equation}
\begin{equation}
\int d\mu_{GH} = 1.
\label{47}
\end{equation}

In order to calculate $\tilde{G}$ we repeat the rescaling of the
Schr\"odinger equation using (\ref{21}), (\ref{44}) and (\ref{37}) and
get
\begin{equation}
\left[ - \frac12 \frac{\partial^2}{\partial x^2} + \frac18 \left(
\frac{\partial}{\partial x} G(x;\{\zeta\})\right)^2 \right]\chi_0(x)
= \tilde{\epsilon}_1 \chi_0(x)
\label{48}
\end{equation}
and
\begin{equation}
\chi_0(x) = \exp \{- \frac12 \tilde{G}(x;\{\zeta\})\}.
\label{49}
\end{equation}
It follows for $x \to + \infty$
\begin{eqnarray}
\tilde{G}(x;\{\zeta\}) &=& \epsilon G(x;\{\zeta\}) \nonumber \\
& & + \log (\epsilon \frac{\partial}{\partial x} G(x;\{\zeta\}) \nonumber \\
& & + O(1).
\label{50}
\end{eqnarray}

\noindent
4. Now we consider an example: the singularity $A_2$. The Airy function
is the proper one
\begin{eqnarray}
\pi Bi(\zeta) &=& \int^{\infty}_0 dx \left\{\exp (- \frac13 x^3 + \zeta x)
+ \sin (+ \frac13 x^3 + \zeta x)\right\} \nonumber \\
&=& \int_C dx \exp (- \frac13 x^3 + \zeta x)
\label{51}
\end{eqnarray}
where $C$ is defined as follows. Let $C_q, \; q \in \bbbq$, denote the contour
along the ray
$\{ r e^{2 \pi i q}, \; 0 \leq r < \infty \}$ oriented from zero to infinity.
Then
\begin{equation} \label{a}
C = C_0 - \frac{1}{2} (C_{\frac23} + C_{\frac43}).
\end{equation}
How can this Airy function be used to calculate expectation values? Consider
a polynomial
\begin{equation}
P_M(x) = \sum^M_{r=0} a_r x^r.
\label{52}
\end{equation}
It is natural to define then
\begin{equation}
\langle P_M(x)\rangle = Bi(\zeta)^{-1} P_M \left( \frac{d}{d\zeta}
\right) Bi(\zeta).
\label{53}
\end{equation}
In order that a probabilistic interpretation is possible, the matrix
${\cal P}_M(\zeta)$
\begin{equation}
{\cal P}_M(\zeta) = \left( \begin{array}{ccccc}
1 & \langle x\rangle & \langle x^2 \rangle &... & \langle x^M \rangle
\\
\langle x\rangle & \langle x^2 \rangle & \langle x^3 \rangle & ...&
\langle x^{M+1} \rangle \\[0.5cm]
\langle x^M\rangle & & & &  \langle x^{2M} \rangle
\end{array} \right)
\label{54}
\end{equation}
must be positive (for some $\zeta_M$ and all $M$) at least for
\begin{equation}
\zeta > \zeta_M.
\label{55}
\end{equation}
 From the asymptotic expansion of the Airy function (\cite{8}, equ. 10.4.63)
follows
\begin{eqnarray}
\langle(P_M(x))^2\rangle &=& (P_M(\zeta^{\frac12}))^2 \mbox{ + lower order
terms} \nonumber \\
&\zeta \to \infty&
\label{56}
\end{eqnarray}
so that ${\cal P}_M(\zeta)$ has one positive eigenvalue for large $\zeta$. It
can be shown that the other eigenvalues are positive for large $\zeta$, too.
With this knowledge it suffices to calculate
\begin{equation}
D_M(\zeta) = \det {\cal P}_M(\zeta).
\label{57}
\end{equation}
For low
$M$ we find (for $\zeta \to \infty$) e.g.
\begin{equation}
D_1(\zeta) = \frac{1}{2\zeta^{\frac12}} + O(\zeta^{-2})
\label{58}
\end{equation}
\begin{equation}
D_2(\zeta) = \frac{1}{4\zeta^{\frac32}} + O(\zeta^{-3}).
\label{59}
\end{equation}
Assume that $D_M(\zeta) > 0$ for $\zeta \to \infty$ has been shown. Then
$\zeta_M$ is the largest zero of $D_M$. For $M = 1$ we obtain (using the
tables in \cite{8})
\begin{equation}
\zeta_1 = 0.4003.
\label{60}
\end{equation}
Finally we have to prove
\begin{equation}
\zeta_c = \sup_M \zeta_M < \infty
\label{61}
\end{equation}
which is so far only wishful thinking.

In the Greensite-Halpern approach we have to solve
\begin{equation}
\left[ - \frac12 \frac{\partial}{\partial x^2} + \frac18
(x^2-\zeta)^2\right] \chi_0(x) = \tilde{\epsilon}_1\chi_0(x).
\label{62}
\end{equation}
By symmetry we have
\begin{equation}
\langle x^{2n+1}\rangle_{GH} = 0, \mbox{ all n}.
\label{63}
\end{equation}
Moreover we find e.g.
\begin{equation}
\langle x^2\rangle_{GH} = \left\{ \begin{array}{l}\zeta + \mbox{ lower order
terms for }
\zeta \to \infty,\\
O(|\zeta|^{\frac12}) \mbox{ for } \zeta \to -\infty \end{array} \right.
\label{64}
\end{equation}
as compared with
\begin{equation}
\langle x^2\rangle = \zeta
\label{65}
\end{equation}
from Airy's differential equation (\cite{8}, eqn. 10.4.1). The
difference between the two approaches becomes more striking if we compare
the dispersions
\begin{equation}
\langle x^2 \rangle - \langle x \rangle^2 = D_1(\zeta)
\label{66}
\end{equation}
\begin{equation}
\langle x^2 \rangle_{GH} - (\langle x \rangle_{GH})^2 = \langle x^2
\rangle_{GH}.
\label{67}
\end{equation}

\end{document}